\documentclass[aps,nofootinbib,superscriptaddress,twocolumn,preprintnumbers]{revtex4}
\usepackage{graphicx,amsmath,amssymb,amsbsy,latexsym,verbatim}
\usepackage{mathpazo}
\usepackage{undertilde}
\usepackage[english]{babel}

\newcommand{\f}{\frac}

\newcommand{\dd}{\mathrm{d}}
\def\tr{{\rm Tr \,}}
\def\R{\mathbb{R}}
\def\H{\mathcal{H}}

\def\id{\mathbb{1}}
\def\K{\kappa}
\def\lin{\ell}

\def\lp{{\ell}_{\rm Pl}}

\newcommand{\be}{\nopagebreak[3]\begin{equation}}
\newcommand{\ee}{\end{equation}}
\newcommand{\ba}{\nopagebreak[3]\begin{eqnarray}}
\newcommand{\ea}{\end{eqnarray}}
\newcommand{\nn}{\nonumber \\}
\newcommand{\bc}{\begin{comment}}
\newcommand{\ec}{\end{comment}}

\newcommand{\bra}[1]{\ensuremath{\,\langle\,#1\,|\,}}
\newcommand{\ket}[1]{\ensuremath{\,|\,#1\,\rangle\,}}
\newcommand{\ip}[2]{{\,\langle\,#1\,|\,#2\,\rangle\,}}
\newcommand{\den}[2]{{\,|\,#1\,\rangle\langle\,#2\,|\,}}

\newcommand{\bk}[3]{{\langle\,#1\,|#2|\,#3\,\rangle\,}\!}

\begin{document}

\title{Single particle in quantum gravity \\ and Braunstein-Ghosh-Severini entropy of a spin network}   

     \author{Carlo Rovelli}      
     \email{rovelli@cpt.univ-mrs.fr}         
     \affiliation{Centre de Physique Th\'eorique de Luminy\footnote{Unit\'e mixte de recherche (UMR 6207) du CNRS et des Universit\'es de Provence (Aix-Marseille I), de la M\'editerran\'ee (Aix-Marseille II) et du Sud (Toulon-Var); laboratoire affili\'e \`a la FRUMAM (FR 2291).}
, Case 907, F-13288 Marseille, EU}     
     \author{Francesca Vidotto}
     \email{vidotto@cpt.univ-mrs.fr}
     \affiliation{Centre de Physique Th\'eorique de Luminy\footnote{Unit\'e mixte de recherche (UMR 6207) du CNRS et des Universit\'es de Provence (Aix-Marseille I), de la M\'editerran\'ee (Aix-Marseille II) et du Sud (Toulon-Var); laboratoire affili\'e \`a la FRUMAM (FR 2291).}
, Case 907, F-13288 Marseille, EU}
     \affiliation{Dipartimento di Fisica Nucleare e Teorica,
        Universit\`a degli Studi di Pavia \\ and Istituto Nazionale
        di Fisica Nucleare, Sezione di Pavia, via A. Bassi 6,
        27100 Pavia, EU}
\date{\today}

\begin{abstract}
Passerini and Severini have recently shown that the Braunstein-Ghosh-Severini (BGS) entropy $S_\Gamma=-\tr[\rho_\Gamma \, \log\rho_\Gamma]$ of a certain density matrix $\rho_\Gamma$ naturally associated to a graph $\Gamma$, is maximized, among all graphs with a fixed number of links and nodes, by regular graphs.   We ask if this result can play a role in quantum gravity, and be related to the apparent regularity of the physical geometry of space.   We show that in Loop Quantum Gravity the matrix $\rho_\Gamma$ is precisely the Hamiltonian operator (suitably normalized) of a non-relativistic quantum particle interacting with the quantum gravitational field, if we restrict elementary area and volume eigenvalues to a fixed value.  This operator provides a spectral characterization of the physical geometry, and can be interpreted as a state describing the spectral information about the geometry available when geometry is measured by its physical interaction with matter.  It is then tempting to interpret its BGS entropy $S_\Gamma$ as a genuine physical entropy: we discuss the appeal and the difficulties of this interpretation. 
\end{abstract}

\maketitle

\section{Introduction}

The content of this paper is twofold. First, we construct a theory describing a single quantum particle interacting with the quantum gravitational field, in the approximation where the particle moves slowly. We focus on the quantum dynamics of the particle, and study how this is affected by the quantum discreteness of spacetime. We use Loop Quantum Gravity (LQG) canonical methods \cite{Rovelli,Thiemann,Ashtekar:2004eh}.  This simple model might be of help in view of the general problem of describing the behavior of matter on a quantum spacetime. But it can also teach us how the quantum properties of geometry can be measured by a material apparatus.   Interestingly, the Hilbert space of the particle turns out to be naturally dependent on the geometry, and to be restricted to states with support on nodes of the spin network that describes the state of spacetime. 
Our main result is the construction of the Hamiltonian operator that governs the particle dynamics.  The operator  depends on the state of the quantum gravitational field, since the particle energy is both a function of the particle degrees of freedom and of the geometry. 

The second part of the paper is based on the observation that in the approximation where we can take the elementary area and volume quantum numbers of the spin-network describing the quantum gravitational field all equal, the Hamiltonian of the particle, suitably normalized, is the Braunstein-Ghosh-Severini (BGS) density matrix $\rho_\Gamma$ of the graph $\Gamma$ of the spin network describing the quantum spacetime \cite{Braunstein:2004}.  In the context of graph theory, Braunstein, Ghosh and Severini define  a ``von-Neumann entropy" $S_\Gamma=-Tr[\rho_\Gamma \, \log\rho_\Gamma]$ associated to a graph $\Gamma$. In a recent paper \cite{Passerini:2008}, Passerini and Severini show that among all graphs with a given number of nodes and links, this entropy is maximized by \emph{regular graphs}: graphs where the links are most uniformly distributed among the nodes.  This implies that --in the approximation given--, the BGS entropy of the (normalized) Hamiltonian is maximized by spin networks with a uniform graph. In other words, there exists a notion of ``entropy" associated to quantum states of space, which is maximized by uniform states.  It is natural to ask if this graph-theory result can be used in the context of quantum gravity, and if it has a physical interpretation. It is even tempting to try to use the Passerini-Severini theorem, or some extension of the same, to justify the apparent uniformity of physical space as it is observed via its interaction with matter.  We discuss this possibility, and its difficulties, in the final part of the paper. In particular, we argue that the BGS entropy cannot be directly taken as a measure of gravitation entropy, but it may indirectly indicate a possible definition of such entropy. 

\section{Particle on a quantum gravitational field}

\subsection{General Relativity with a particle}
In this section we construct the quantum theory of General Relativity interacting with a single particle. In the conventional (ADM) framework, the canonical phase space of this theory is coordinatized by $(\,q_{ab}(   x),\,\pi^{ab}(   x),\,    X^a,\,    P_a\,)$
where $q_{ab}(   x)$ is 3d metric 
(\,$a,b=1,2,3$\,)
with conjugate momentum $\pi^{ab}(   x)$ and
$(   X^a,\,    P_a\,)$ are the position and the canonical momentum of the particle.
$x=\{x^a\}$ denotes the coordinates of a 3d constant-time surface.
The dynamics is given by a Hamiltonian constraint 
$C(   x)= H_{\!\!A\!D\!M}(   x) + \delta^3(   x,   X) P_{0}$
where $H_{\!\!A\!D\!M}$ is the Hamiltonian constraint of pure gravity and $P_{0}$
is the energy of the particle. 

We take the approximation in which the velocity of the particle is small%
\footnote{This does not imply that we take $c\to\infty$ in the theory \cite{Rovelli:2008cj}.
It only means that we make measurements in frames where the particle moves slowly.}
that is we take $ P^2\!\ll\! m^2 $ and $P_{0} = \sqrt{   P^2 + m^2} \sim m + \f{   P^2}{2m}\,$. In this approximation, 
\be 
C(   x)= H_{\!\!A\!D\!M}(   x) + 
\delta^3(   x,   X)m + 
\delta^3(   x,   X)\f{   P^2}{2m}
\ .
\ee
We focus on the last term, the Hamiltonian of the particle on the given gravitational field.  Here 
$P^2 = q^{ab}(x)P_{a}P_{b}\,$; notice the presence of the metric tensor, which gives the interaction between the particle and the gravitational field. Smeared with the Lapse function $N(x)$ the Hamiltonian reads
\ba 
H
&=&\f{1}{2m} 
\int\dd x \,N(x)\ \delta^3(x,X)\ q^{ab}(x)\ P_{a}\ P_{b} 
\nn &=&\f{ N(X)}{2m}\ q^{ab}(X)\ P_{a}\ P_{b} 
\ .
\ea
In what follows we take $N(x)=1$ .

In view of the quantization it is convenient to rewrite the theory in terms of Ashtekar variables.  We shift to the variables
$(\,E^{ai}(   x),\, A^i_a(   x),\,    X^a,\,    P_a\,)$ 
where $A^i_a(   x)$ is the Ashtekar$-$Barbero connection \cite{Ashtekar:2004eh},
and $E^{ai}(   x)$ is the corresponding conjugate momentum;
this is the densitized triad,
related to the 3d metric by $E^{ai}E^{bi}= q q^{ab}$,
with $i=1,2,3$, where $q=\det q_{ab}$.
In terms of these variables, the Hamiltonian of the particle reads
\be  \label{hamcla}
H=  \f1{2m} \,
\f{E^{ai}(X)E^{bi}(X)}{{q(X)}}\,
  P_{a}P_{b} 
\ .
\ee
This is the Hamiltonian 
that we want to define in the quantum theory.

\subsection{Quantum theory} %$^{}$\\ \centering 
The state space  ${\cal H}={\cal H}_{\scriptscriptstyle\rm L\!Q\!G} \!\otimes\! \H_{\scriptscriptstyle\rm P}$ of the quantum theory will be the tensor product
of the gravity state space ${\cal H}_{\scriptscriptstyle\rm L\!Q\!G} $ and a space of particle states  $ \H_{\scriptscriptstyle\rm P}$. 
In the first, we chose a spin network basis that diagonalizes areas and 
volumes.  That is, we take
the LQG state space ${\cal H}_{\scriptscriptstyle\rm L\!Q\!G}$, 
spanned by a basis $\ket s$ labeled by spin network $s$.
A spin network $s=(\Gamma, j_\lin, \nu_n)$ is a graph $\Gamma$ 
immersed in $\R^3$,  colored with quantum numbers $j_{\lin}$
on each link $\lin$ and quantum numbers $\nu_{n}$ on each node $n$.
Here $\nu_{n}$ are volume eigenstates.  That is, we choose a 
basis $\ket{\nu_n}$ in the space of the intertwines at each node, 
that diagonalizes the volume of the node. These states are linear combinations 
of states spanned by the conventional intertwiner basis. The states  $\ket s$  are orthonormal.

In the particle state space $\H_{\scriptscriptstyle\rm P}$ we chose a (possibly generalized) basis $\ket x$ that diagonalizes the particle's position $X$. We write states forming a basis in ${\cal H}={\cal H}_{\scriptscriptstyle\rm L\!Q\!G} \!\otimes\! \H_{\scriptscriptstyle\rm P}$ in the form
\be 
\ket{s,x}
\!
\equiv
\!
\ket s \!\otimes\! \ket x  
\subset 
{\cal H}_{\scriptscriptstyle\rm L\!Q\!G} \!\otimes\! \H_{\scriptscriptstyle\rm P} \ .
\ee

The definition of the appropriate scalar product on $\H_{\scriptscriptstyle\rm P}$ requires some care.  
In usual non-relativistic wave mechanics, the scalar product is usually written in the form
\be 
      \ip\psi\phi =  \int\dd x ~ \overline\psi(x)\phi(x) \, ,
\ee
but in this expression the coordinates $x$ are the Cartesian coordinates of space, that contain metric information. In general coordinates, the above expression reads 
\be 
      \ip\psi\phi =  \int\dd x\sqrt{q}\  \ \overline\psi(x)\phi(x) \, .
\ee
Taking this as the definition of the states $\ket x$
is equivalent to write in $\H_{\scriptscriptstyle\rm P}$ the resolution of the identity 
\be\label{id}
 \id_{\scriptscriptstyle\rm P}  = \int\dd x~\sqrt q\den xx,
 \ee
that is, to have states normalized as 
\be
 \ip xy  = \f1{\sqrt{q(x)}}\ \delta(x,y). 
 \ee
Equation \eqref{id} is the appropriate resolution of the identity in $\H_{\scriptscriptstyle\rm P}$. However, in the theory we are interested in,  $\sqrt q$ is an \emph{operator} on 
${\cal H}_{\scriptscriptstyle\rm L\!Q\!G}$. Therefore the scalar product in $\H_{\scriptscriptstyle\rm P}$ depends on 
${\cal H}_{\scriptscriptstyle\rm L\!Q\!G}$.%
\footnote{That is, by itself $\H_{\scriptscriptstyle\rm P}$ is a vector space but not a Hilbert space. It is ${\cal H}={\cal H}_{\scriptscriptstyle\rm L\!Q\!G} \!\otimes\! \H_{\scriptscriptstyle\rm P}$ which is a Hilbert space.} 
 On  ${\cal H}={\cal H}_{\scriptscriptstyle\rm L\!Q\!G} \!\otimes\! \H_{\scriptscriptstyle\rm P}$ we can sandwich the right hand side of (\ref{id}) between two states $\bra s$ and $\ket s$ in ${\cal H}_{\scriptscriptstyle\rm L\!Q\!G}$. In this way, we have that the operator $\sqrt q$ is replaced by its expectation value in $\ket s$:
\be
\bk s{\int\!\dd x~\sqrt q\den xx}s  =
\int\dd x~ \bk s{\!\sqrt q}s\den xx .
\ee
But the volume operator vanishes everywhere except at the nodes:
\be
{\sqrt{q(x)}}\ket s = \sum_{n\in \tilde N(s)}\nu_n\ \delta(x,x_n)\ket s, 
\ee
where we have indicated as $\tilde N(s)$ the set of the nodes $n$ of     $s$ that have nonvanishing volume eigenvalue $\nu_n$. \mbox{Using} the last equation in the previous one gives
\ba
\bk{s}{ \id_{\scriptscriptstyle\rm P}}{s} &=&
\sum_{n\in \tilde N(s)}\nu_n\den{x_n}{x_n}, 
 \ea
This gives the resolution of the identity in ${\cal H}_{\scriptscriptstyle\rm L\!Q\!G} \!\otimes\! \H_{\scriptscriptstyle\rm P}$
\be
 \id =\sum_s \sum_{n\in \tilde N(s)}\nu_n\den{s,x_n}{s,x_n}.
 \ee
 It follows that all wavefunctions $\psi(x)=\ip x\psi$ that have the same values  $\psi(x_n)=\ip{x_n}\psi$ on the nodes are to be identified as elements of the \emph{Hilbert} space of the theory.   The relevant Hilbert space is spanned by the states $\ket{s,x_n}$, where $n\in \tilde N(s)$, satisfying 
\be
 \ip{s,x_n}{s',x_{n'}} =\f1{\nu_n}\delta_{ss'}\delta_{nn'}.
 \ee
The quantum states of the particle cannot be considered independently from those of the geometry. 
Remarkably, the same conclusion has been reached in \cite{Ashtekar:2009mb}, in the context of quantum cosmology, from a rather different set of arguments. 

For later purposes, it is also convenient to define the states 
\be
\ket{\utilde{x}} := \f 1{\sqrt[4]{q(x)}} \ket{x}
\ee
that satisfy 
\ba \label{id2}
 \id_{\scriptscriptstyle\rm P}  = \int\dd x~ q\den{\utilde{x}}{\utilde{x}} &=& \sum_{n\in \tilde N(s)}\nu^2_n 
 \den{\utilde{x_n}}{\utilde{x_n}} ,
\\
 \label{braket}
 \ip{s, \utilde{x_n}}{s', \utilde{x_{n'}}} &=&\f1{\nu_n^2}\ \delta_{ss'}\delta_{nn'}
 \ea
and the orthonormal states $\ket{\tilde{x}} := \sqrt[4]{q(x)} \ket{x}$.
The definition of these other bases is only for convenience of notation and is not strictly needed; in particular, it is not meant to conceal $1/\nu$ factors, since inverse volume factors are not harmful due to our definition of the Hilbert space.  This completes the definition of the Hilbert space of the theory. 

The quantum theory is obtained by promoting the phase space variables to operators on
this Hilbert spaces. In particular the particle momentum operator is
$
P_{a}Ê= -i\hbar \ D_a %\p{}{x^a}
 \ ,  $ being $D_a$ the covariant derivative,
and the densitized triad operator is defined%
\footnote{In the recent literature, 
this operator is mainly used 
as a flux operator, that is, smeared with a two surface. As we shall see below, here the space index $a$ 
turns out to be contracted with a space derivative rather than with 
the normal of a surface; hence we need the original form of the
operator.   The expression \eqref{E} is directly obtained from the connection representation
acting with $E^{ai}(x)=\K\hbar \f\delta{\delta A_a^i(x)}$ on the 
holonomy $h_{\lin}(A)={\cal P}\exp{\int_{\lin} \!\dd t \,\dot\lin^a(t) A^i_a(\lin(t))\tau^i}$, which is the kernel of the integral transform relating 
the loop and the connection representations.}  by
\cite{Rovelli:1989za,Rovelli}
\be \label{E}
 %\hat
 E^{ai}(x) \ket s = \K\hbar \sum_{\lin} \int_{\lin} \!\dd t \,\dot\lin^a(t)\,\delta^3(x,\lin(t))\ket{s,\tau^{i}} 
 \ee
where $\dot \lin^a(t)$ is the tangent to the link in the point $t$, 
the sum is over the links $\lin$ of the spin network $s$ 
and $\ket{s,\tau_{i}}$ indicates the spin network $\ket{s}$ 
with a grasp in the point $\lin(t)$. 
Here $\K:=8\pi G_{\rm Newton}$. We also recall for later convenience that 
$\ket{s,\tau^{i}\tau^{i}}= j_\lin ( j_\lin +1)  \ket s  $ if the two grasps act on the same point, situated on the link $\lin$.

\subsection{The Hamiltonian operator}

We now study the quantum operator in $\cal H$ corresponding to the phase space function  \eqref{hamcla}.   We chose the following symmetric ordering of this operator
\be  \label{hammasq}
H=  \f1{2m} \,\f1{\sqrt q}P_a
{E^{ai}(X)E^{bi}(X)}
P_{b}  \,\f1{\sqrt q}
\ .
\ee
Sandwiching this operator between two resolutions of the identity \eqref{id} gives 
\ba 
H\!&=&\!  \f1{2m}  \int \dd x\, \dd y    \den xx   \,P_a
{E^{ai}(X)E^{bi}(X)}
P_{b}  \, \den yy \nn
&=& \! -\f{\hbar^2}{2m}  \int \dd x\, \dd y \ket x   D_a
{E^{ai}(x)\f{\delta(x,y)}{\sqrt q}E^{bi}(y)}
D_{b}  \bra y 
\ .\nonumber
\ea
The product of field operators $E^{ai}(x)\delta(x,y)E^{bi}(y)$ at the same point is ill defined. This is to be expected and we can interpret it as a manifestation of the infinities due to the
the pointlike nature of the particle.  To deal with these divergences, we point-split the operators by 
regularizing the delta distribution. We replace it with a smearing function $f_{R}(x,X)$ where  $ f_{R}$ is zero if $|x-X| \geq R$
and $ f_{R}=\f1{V_R}=\f3{4\pi R^3}$ if $|x-X| \leq R$.
Here $R$ is a small real number, representing a regularization ($R$ can be interpreted as the coordinate size of the particle), and the distance $|x-X|$ is the coordinate distance (equivalently: the distance in a fiducial background metric $q^{0}_{ab}$, which does not affect the final result).
This gives the regularized Hamiltonian 
\be  \label{hammasq2}
\bk xHy = 
 -\f{\hbar^2}{2m}  \f{D_a
E^{ai}(x)f_R(x,y)E^{bi}(y)
D_{b}}{{q^{\f34}(x)}{{q^{\f34}(y)}}}
\ .
\ee
or
\be  \label{hammasq2}
\bk{\utilde{x}}H{\utilde{y}} = 
 -\f{\hbar^2}{2m}  \f{D_a
E^{ai}(x)f_R(x,y)E^{bi}(y)
D_{b}}{{q(x)}{{q(y)}}}
\ .
\ee

Let us now fix a spin network $s$ and study the matrix elements of this operator between two states $\ket{s,\psi}$ and $\ket{s,\phi}$, where $\psi,\phi\in  \H_{\scriptscriptstyle\rm P}$
 \be
  \bk{\psi}{H_s}{\phi} \equiv  \bk{s,\psi}{\!H\!}{s,\phi}.
\ee
Inserting two resolutions of the identity \eqref{id2}
 \ba
 \bk{\psi}{H_s}{\phi} &=& 
  \bk{s,\psi}{\id_{\scriptscriptstyle\rm P}H\;\id_{\scriptscriptstyle\rm P}}{s,\phi} 
\\ &=& \nonumber \int \dd x\ \dd y 
 \bk{s,\psi}{ q \den{\utilde{x}}{\utilde{x}} H  \den{\utilde{y}}{\utilde{y}} q}{s,\phi}
 \ea
Inserting the Hamiltonian \eqref{hammasq2}  into this expression gives
 \ba
  \bk{\psi}{H_s}{\phi}  &=&-\f{\hbar^2}{2m} \int \dd x\ \dd y \ \ 
 \ip{s,\psi}{\utilde{x}}
\nn
 && \times 
  D_a
E^{ai}(x)f_R(x,y)E^{bi}(y)
D_{b}
  \ip{\utilde y}{s,\phi}\nn 
  &=&\f{\hbar^2}{2m} \int {\dd x}\ {\dd y}\ \ \partial_a\overline{\psi(\utilde x)}
 \ {\partial_{b}\phi(\utilde y)} \ f_R(x,y)
 \nn &&\hspace{3em} \times  
\bk{s}{
E^{ai}(x)E^{bi}(y)}{s}
\ea
where $\psi(\utilde x)=\ip{\utilde x}\psi$ and 
we have integrated by parts. The integration by parts does not give
a boundary contribution as we can assume the wave function of the 
particle to vanish at infinity (or, more simply, we can take the space to
be compact.) Inserting the explicit expression \eqref{E} for the 
operators $E^{ai}(x)$ into this expression gives
\ba
 \bk{\psi}{H_s}{\phi}\!
 &=&
 \f{\hbar^2}{2m}(\K\hbar)^2  \!\!\!
\int\!{\dd x}\  {\dd y}\ ~ {\overline{ \partial_{a}\psi}(\utilde x)}
 \partial_{b} \phi(\utilde x)
 f_{R}(x,y) 
\nn 
&& \!\!\!\!\! \times \
 \sum_{\lin\lin'} \!\int_{\lin} \!\dd s \ \dot\lin^a(s)\,\delta(x,\lin(s)) 
\nn && \!\!\!\!\! \times \
\!\!\int_{\lin'}\!\dd t\, \dot\lin^b(t)\,\delta(y,\lin'(t)\!) 
 \ip {s}{s , \tau^i_{\lin} \tau^i_{\lin'}}\nn
 &=&
           \f{\K^2\hbar^4}{2m}  
           \sum_{\lin\lin'}  
           \int_\lin\! \!   \dd s    \int_{\lin}\!       \dd t ~ 
           \overline{ \partial_{s}\psi(\utilde{\lin(s)}\!)} \,  \partial_{t} \phi(\utilde{\lin'(t)}\!)
 \nn &&\ \times \
            f_{R}(\lin(s),\lin(t)\!)  \
            \ip {s}{s , \tau^i_{\lin} \tau^i_{\lin'}}
\ea            
where we have used
$\dot \lin^a(s)\partial_a \psi(x)=\partial_s\psi(\lin(s))$. 
In the limit where $R$ is small,
the only contribution to this sum
comes when $\lin=\lin'$ or when 
$\lin$ and $\lin'$ meet on a node.
This second case however gives
a lower order contribution to the sum.
Furthermore, in the same limit
$
\ket{s, \tau^i_{\lin} \tau^i_{\lin'}}
=  j_\lin ( j_\lin +1) \ket s %\delta_{\lin\lin'}
$. Bringing all this together we have
\ba
  \bk{\psi}{H_s}{\phi}
 &=& 
           \f{\K^2\hbar^4}{2m}  \ \ \ 
           \sum_{\lin} \   j_\lin ( j_\lin +1)
            \int_\lin\!   \dd s    \int_{\lin}\!       \dd t ~ 
            \nn &&\!\!\!\!\! \times\ 
           \overline{ \partial_{s}\psi(\utilde{\lin(s)}\!)} \,  \partial_{t} \phi(\utilde{\lin(t)}\!)
\             f_{R}(\lin(s),\lin(t)\!).
\ea   
   
It is time to deal with our regularization.  To this aim, recall that
the Hilbert structure of the state space of the particle depends only
on the value of the wave function on the nodes. The value
of the wave function along a link that connects two nodes is physically
irrelevant. Consistently, the expression $\int_\lin\!   \dd s \  \partial_{s} {\psi}(\lin(s)\!)$
contained in the equation above suggests that this equation should only depend on the difference  
\be
\Delta_{\lin} \psi := \int_\lin\!   \dd s \  \partial_{s} {\psi}(\utilde{\lin(s)}\!)={\psi}(\utilde{\lin_f})-{\psi}(\utilde{\lin_i}),
\ee
where 
$\lin_f$ and $\lin_i$ are the initial and final points of the link $\lin$.  This is the case for large $R$, but not for small $R$, where the regularization function $f_R(\lin(s),\lin(t)\!)$ cuts the integrals off.  However, we can exploit the freedom in choosing the wave function $\psi$ to go to a limit where the overall integral does not depend on the regulator. This can be done, for each $R$, by choosing $\psi(x)$ to be constant around each node, so that the region along the link where it varies is always smaller than $R$.  In the limit, 
$\partial_{s} {\psi}(\lin(s))\to \Delta_{\lin} \psi\, \delta(s,s_0)$, where $s_0$ is an arbitrary point on the link. Doing so, we have in the limit
\be
  \bk{\psi}{H_s}{\phi}=
           \f{\K^2\hbar^4}{2mV_R}  \ \ \ 
           \sum_{\lin} \   j_\lin ( j_\lin +1)\ \overline{\Delta_{\lin} \psi} \ \Delta_{\lin} \phi. 
          \ee
We absorb the dimensionless divergent factor $V_R\to\infty$ into a renormalization of the mass\footnote{By this we simply mean that we absorb the infinity into (any) one of the constants. An infinite factor is to be expected in this context, because the particle Hamiltonian we started from is ill defined on a discrete geometry, due to the product of the triad operators at a point.  Physically, of course, we can say that it is the limit that defines the continuous derivative that must be replaced by a finite difference, on a discrete geometry.}, defining $m^*=m V_R$, and, using also $\lp^2=\hbar G_{Newton}$, we have finally
\be \label{den}  
       \bk{\psi}{H_s}{\phi} =
      \f{(8\pi\hbar)^2\lp^4}{2m^*} 
      \sum_{\lin} \   j_\lin ( j_\lin +1)\ \overline{\Delta_{\lin} \psi} \ \Delta_{\lin} \phi \, .
\ee
Equivalently, if we define the particle states 
\be
\ket{\utilde\lin} :=    \ket{\utilde{\lin_f}}\!-\!\ket{\utilde{\lin_i}}\!
\ee
we can write  the Hamiltonian in the total Hilbert space ${\cal H}$ in the form
\be \label{den}  
      H =
      \f{(8\pi\hbar)^2\lp^4}{2m^*} \ \
      \sum_{s,\lin\in s}\ \  j_\lin ( j_\lin +1)\ 
      \den{s,\utilde\lin}{s,\utilde\lin}
\, .
\ee
This remarkably simple operator is the Hamiltonian operator of the particle on a gravitational field, and the main result of this first part of the paper.  We understand that the derivation above is somewhat acrobatic and the regularization a bit brutal.  But the final form of the operator is simple, convincing, and has the required properties, including the correct low energy limit.  In the appendix, we provide a simpler and more heuristic alternative derivation of the same operator. 

Let us see some properties of this operator.  First, the dimension of the first factor is 
$
\left[{(8\pi\hbar)^2\lp^4}/{2m^*}
 \right]
 =$
[Energy]$\times$[Length]${}^6$; 
the states $\ket{\utilde{x_n}}$ have dimension of \ 
[Length]$^{-3}$ from \eqref{braket}.
Thus $H$ is indeed an energy.

Second, $H$ is positive semi-definite. This can be shown by noticing that
it is a linear combination  of projector operators with positive coefficients. Therefore no state can have a negative expectation value. 

Next, the ground state of $H$ corresponds to the case
where the particle is maximally delocalized.
In analogy with the vacuum state of free particle, let us define the ground state 
$\psi_{0}(\utilde x_n)=\ip{\utilde{x_n}}{\psi_{0}} = c_{0}$ constant
over the $\ket{{\utilde{x_n}}}$ states. (That is, its value doesn't depend on $n$). 
Then $\ip{\utilde{\lin}}{\psi_{0}}=0$ and therefore $H\ket{\psi_0}=0$.
However, if the particle is in this ground state, the probability of finding it in different nodes is not the same for all nodes, as the states $\ket{\utilde{x_n}}$ are not normalized.  This probability is rather
\be
P_n = \f{|\ip{\utilde{x_n}}{\psi_0}|^2}{\ip{\utilde{x_n}}{\utilde{x_n}}\ip{\psi_0}{\psi_0}}= \f{v^2_n}{\sum_m \! v^2_m}.
\ee
That is, in the ground state the particle is more likely to be found on nodes with a larger volume. 

It is interesting to notice that if we consider a 
state entirely localized on a single node $n$, the 
expectation value of its energy turns out to be 
\ba \label{mven}
\bk {s, \tilde{x_n}}{\!H\!}{s, \tilde{x_n}}
&=&
\f{(8\pi\hbar)^2 \lp^4 }{2m^*} \sum_{{\lin}\in n}  
j_\lin ( j_\lin +1)
\,|\! \ip{s, \utilde{x_n}}{s, \tilde{x_n}}\! |^2
\nn &=&  
\f{(8\pi\hbar)^2 \lp^4 }{2m^* \nu^2_n}\ \sum_{{\lin}\in n}
 j_\lin ( j_\lin +1).
\ea 
If we assume that the volume of the node has Planck scale, 
 $\nu_{n}\approx\lp^3$, 
 then $\bk {s, \tilde{x_n}}{\!H\!}{s, \tilde{x_n}}\approx 
\f{\hbar}{2m^*G_{\rm Newton}}= \f\hbar{r_{S}}$, i.e.\ the energy is given by Planck constant over the Schwarzschild radius of the particle $r_{S}$.  
 
Finally, we observe that the form \eqref{den} of the Hamiltonian operator that we have found is not too surprising. As we show in the Appendix, there is a simple minded way of obtaining it, starting from the discretization of the Schr\"odinger Laplacian of the particle on a lattice, and writing the lattice geometry in terms of areas and volumes.  
 
\section{BGS entropy of gravity}

As anticipated in the introduction, the Energy operator \eqref{den} that we have found is strictly related to the BGS density matrix of the graph of the spin network, and its BGS entropy is maximized by regular graphs.  In this section, we begin by recalling some basic facts of BGS graph theory, we illustrate the mathematical relation between this theory and LQG and we investigate if the BGS results can be interpreted physically. 

\subsection{BGS entropy of a graph}

We begin by reviewing the notion of entropy on a graph as introduced in \cite{Braunstein:2004}.
A graph $\Gamma=\{N,L\}$ is defined just as a set of nodes $N(\Gamma)=\{1,2,...,n\}$ and a set of couples of nodes $L(\Gamma)$, namely links connecting nodes. %
%\footnote{
The graphs under consideration are said to be simple and undirected: links are not colored and not oriented, differently with respect to the general case in LQG.%}
The  graph is constructed with the following rules: 
two nodes are connected at most by one link and  a link cannot close
on the same node where it starts. Two nodes are said to be \emph{adjacent} if they are 
connected by a link. One can construct an \emph{adjacency} matrix $A(\Gamma)$ 
whose elements are $[A(\Gamma)]_{n,m}=1$ if $\{n,m\}\in L(\Gamma)$ and
$[A(\Gamma)]_{u,v}=0$, otherwise. The number of links attached to a node $n$ gives
the \emph{degree} of the node $n\in V(\Gamma)$,
denoted by $d_n$. 
The sum of the degree of each nodes is the \emph{degree-sum} of the graph 
$d_{\Gamma}=\sum_{n\in N(\Gamma)}d_n$. 
The \emph{degree} matrix $\Delta(\Gamma)$ is a diagonal matrix,
having on the diagonal the degree $d_n$ correspondent to each node. 
The \emph{combinatorial Laplacian} matrix $L(\Gamma)$
is a positive semidefinite $n\times n$ matrix, defined as the difference
between $\Delta(\Gamma)$ and $A(\Gamma)$. 
Once one divides $L_{\Gamma}$  by the degree-sum $d_\Gamma$
the resulting matrix $\rho_{\Gamma}:=\frac{L(\Gamma)}{d_{\Gamma}}=\frac{L(\Gamma)}{\tr(\Delta(\Gamma))}$
is Hermitian, positive semidefinite and unit-trace, 
and because of these properties
it is said to be the \emph{density} matrix of the graph. 

The \emph{BGS entropy} of the graph $\Gamma$ is defined%
\footnote{Following the conventions of information theory,
we set logarithm to be base 2 \ and \ $0\log0=0$ .}
as the von Neumann entropy of its density matrix: 
\be
S(\Gamma)=-\tr[\rho_\Gamma\log\rho_\Gamma].
\ee

In \cite{Passerini:2008}, Passerini and Severini have shown that at fixed number of nodes and vertices, the BGS entropy is maximized by regular graphs.  Regular means here that the distribution of the degrees of the nodes is uniform. 

Does this result have physical interpretation, in the context of LQG?

\subsection{The Hamiltonian operator codes \\
the information about the geometry}

The properties of a physical system are revealed by the interactions of the system with its surrounding. The geometry of spacetime is a physical system; it interacts with the rest of the world by affecting the energy momentum tensor of matter, or, in the canonical picture, its energy.  Therefore the energy of matter can be seen as an interaction Hamiltonian between matter and gravity. In other words, the only tool we have to detect the spacetime geometry is via its effect on our material measuring devices; this effect is coded in the way the metric tensor enters the matter Hamiltonian. 

In the gravity+particle system considered above, a measurement of the particle's energy measures particle as well as metric properties.  In non-general-relativistic physics, we view the spacetime geometry as given; then the energy is just an observable of the particle.  But this energy is also an observable of the geometry: we get information on the geometry by measuring energy levels and energy eigenstates of matter on this geometry.  

Alain Connes has long stressed the relevance of this \emph{spectral} point of view on the geometry of physical space \cite{Connes}.  To put it pictorially: the only information we have about the universe is via the \emph{spectra} of the light we receive from the sky; or, even more simply: our eyes only see spectra!  The information about the geometry of space which is available to matter is therefore coded precisely into the matter's Hamiltonian operator.  Geometry is nothing else than what determines the Hamiltonian operator of matter. 

From this perspective, the Hamiltonian operator \eqref{den} constructed above codes the information about the gravitational field that is available to the particle. As an operator on $\H_{\scriptscriptstyle\rm P}$, indeed, $H_s:=\bk sHs$ is a function of the state $\ket s$ of gravity.  Different states $\ket s$ determine different operators. This operator can be reconstructed, or ``measured" by a sequence of energy measurements on known particle states.%
\footnote{In general, the diagonal matrix elements of an operator $H$ in a basis $\ket n$ can be simply measured as the expectation value of the basis states: $H_{nn}=\tr[H\den nn]$; while the nondiagonal matrix elements are the difference between the expectation values of  the quantum superposition and the statistical superposition of the corresponding basis states: $H_{nm}=\tr[H\f{\ket n+\ket m}{\sqrt 2}\f{\bra n+\bra m}{\sqrt 2}]-\tr\left[H\left(\f{\den nn}2 +\f{\den mm}2\right)\right]$.}
 From this perspective, the role of operator and state are --in a sense-- reversed: each particle state defines a measurement that allows us to determine (``measure") the operator $H$, and therefore measure the geometry.  
 
 Let us normalize the operator by dividing it by its trace, that is,  define the unit trace operator
\be
     \rho_{\scriptscriptstyle\rm G} :=  H_{s} /\tr H_{s} .
 \ee
The expectation value of this operator%
\footnote{\label{tracciaf} 
In fact, this construction is more general. Consider a general \emph{statistical} state of the geometry, defined by a density matrix $\tilde\rho=\sum_{ss'}(\rho_{\scriptscriptstyle\rm G})_{ss'}\den s{s'}$ on ${\cal H}_{\scriptscriptstyle\rm LQG}$. This as well determines a Hamiltonian operator on $\H_{\scriptscriptstyle\rm P}$
\be\label{Hrho}
       H_{\tilde\rho} = \tr_{\scriptscriptstyle\rm LQG}[H\tilde\rho],
\ee
($\tr_{\scriptscriptstyle\rm LQG}$ is the trace in $\H_{\scriptscriptstyle\rm LQG}$) and a corresponding density matrix on $\H_{\scriptscriptstyle\rm P}$
\be
     \rho_{\scriptscriptstyle\rm G} :=  H_{\rho_{\scriptscriptstyle\rm G}} /\tr H_{\rho_{\scriptscriptstyle\rm G}} .
 \ee  
The map $\tilde\rho\mapsto\rho_{\scriptscriptstyle\rm G}$ so defined sends a density matrix in ${\cal H}_{\scriptscriptstyle\rm LQG}$ to a density matrix in $\H_{\scriptscriptstyle\rm P}$. We get back to the previous case if the $\tilde\rho=\den ss$.}
 on an arbitrary statistical state $\rho_{\scriptscriptstyle\rm P}=\int \dd x\dd x'\rho_{\scriptscriptstyle\rm P}(x,x')\den x{x'}$ on $\H_{\scriptscriptstyle\rm P}$  \emph{of the particle} is 
\be
E =\tr[\rho_{\scriptscriptstyle\rm G}\rho_{\scriptscriptstyle\rm P}].   \label{double}
 \ee
This is an interesting expression. Its signification is clear: it gives the (normalized) mean energy for a certain state of particle and gravity. For a \emph{given} gravitational state, $\rho_{\scriptscriptstyle\rm G}$ can be seen as an operator measuring a property of the particle state: its energy. The other way around, if we assume that we can independently measure  the state of the particle, then each \emph{given} particle state $\rho_{\scriptscriptstyle\rm P}$ can be seen as an operator measuring the ``state" of the gravitational field coded by $\rho_{\scriptscriptstyle\rm G}$. 
The ``density matrix" $\rho_{\scriptscriptstyle\rm G}$ can therefore be interpreted as a representation of the state of the quantum gravitational field; more precisely, of its features that are accessible to the particle via spectral measurements.   

Let us now take the simplifying assumption of disregarding volume and area eigenvalues in the definition of a spin network.%
\footnote{Alternatively, the BGS construction can probably be generalized to include spins and volumes.}  For instance, let $s$ be a spin network with $j_\lin=j$ and $\nu_n=\nu$. Then $\rho_{\scriptscriptstyle\rm G}$ determined solely by its graph $\Gamma$.  And it is easy to see that it is precisely the BGS density matrix of the graph%
 \footnote{Therefore $\rho_\Gamma$ can be written in the form \eqref{den}: this provide a straightforward proof that it is positive semi-definite.}
\be
       \rho_{\scriptscriptstyle\rm G}=\rho_{\Gamma}.
\ee
 
\subsection{The BGS entropy in gravity}

We have seen above that the BGS density matrix of a graph is an expression of a (possibly statistical) state of the geometry, as seen by matter. What is then the BGS entropy of this state?  

First of all, as shown in \cite{Passerini:2008}, this entropy is a quantitative measure of the regularity of the graph.  As such, it is certainly a useful physical tool. But can we push the analogy with statistical mechanics further, and interpret the BGS entropy as a genuine physical entropy? Attempts to define the entropy of the gravitational field are in \cite{Rovelli:1993a, Rovelli:1993b, Connes:1994hv} and \cite{Smolin:1981jt, Smolin:1982jt}, but the  problem of defining this entropy is completely open; for a discussion, see \cite{Rovelli:1993a}.   Can the BGS entropy provide a further tool for dealing with gravitational entropy?  If we could do so, then the Passerini-Severini result that relates this entropy to regularity could have a wide reach, and we could try to use it in order to argue that the observed uniformity of physical space has a statistical origin. The possibility is very tempting.

We have found a structure remarkably similar to that of standard statistical mechanics. First, $\rho_{\scriptscriptstyle\rm G}$ has all the mathematical properties of a statistical density matrix. Second, it has also the interpretation of coding the quantum state of gravity, including, possibly, the statistical quantum state.  We may interpret the trace from $\cal H$ to $\H_{\scriptscriptstyle\rm P}$ that defines $\rho_{\scriptscriptstyle\rm G}$ in \eqref{Hrho} in footnote \ref{tracciaf} as a way of tracing out gravitational degrees of freedom not seen by matter, giving a statistical character to $\rho_{\scriptscriptstyle\rm G}$ even when determined by a pure $\tilde\rho$. Finally, \eqref{double} can be interpreted by saying that if the particle state is known $\rho_{\scriptscriptstyle\rm P}$ is an operator that measures the statistical state $\rho_{\scriptscriptstyle\rm G}$.  Since this structure reproduces that of statistical mechanics, one is tempted to say that $\rho_{\scriptscriptstyle\rm G}$ is a genuine statistical density matrix and therefore its BGS entropy 
\be \label{BGSE}
S_{\rm BGS} = - \tr [\rho_{\scriptscriptstyle\rm G}\log\rho_{\scriptscriptstyle\rm G}]
\ee
is a physical entropy for the gravitational field.    If it looks like a duck, swims like a duck and quacks like a duck, then it probably is a duck.  

But not necessarily.  The problem is that $\rho_{\scriptscriptstyle\rm G}$ is \emph{not} a statistical density matrix.  Its eigenvalues are not probabilities of outcomes. They do not measure numbers of microstates.   They are (normalized) matter energy levels in a certain geometry. Accordingly, we do not see how the BGS entropy \eqref{BGSE} could be interpreted as a physical entropy. There are no arguments that we can see, such as an ergodicity hypothesis, or Bayesian equiprobability prior, motivating the expectation that we should find physical systems maximizing it.  

In quantum mechanics, a pure state $\ket\psi$ determines an observable, $\den\psi\psi$ whose interpretation is to check whether or not the system is in the state $\ket\psi$. A density matrix  $\rho=\sum_i p_i \den{\psi_i}{\psi_i}$ can also be interpreted as an observable, describing the operation of choosing with probability $p_i$ to check whether the system is in the state $\ket{\psi_i}$. The other way around, we can associate to any operator $A$ with unit trace a statistical state $\rho_A$ such that on any state the above operation gives the same result as the mean value of $A$. But the analogy stops here, because the eigenvalues of $A$ are possible measurement outcomes; while the eigenvalues of $\rho$ are probabilities of observing one or another state. In the present case, the eigenvalues of $\rho_{\scriptscriptstyle\rm G}$ are still the possible individual outcomes of the energy measurement, and have no probabilistic interpretation that we can see. 

In fact, one \emph{can} define a density matrix and an entropy based on the Laplacian operator $H$ constructed here, and therefore indirectly on the Laplacian of a graph, which capture the idea of gravitational entropy of the distribution determined by measuring the gravitational field via the energy of a particle. To this purpose, fix the state of the particle to be $\ket{\psi}\in \H_{\scriptscriptstyle\rm P}$, and define the operator 
\be
     H'= \bk{\psi}H{\psi}.
\ee
on ${\cal H}_{\scriptscriptstyle\rm LQG}$. Suppose we measure a certain average value $\bar E$ of the energy for the particle in this state. This determines a probability distribution $\tilde\rho$ for the state of the gravitational field in ${\cal H}_{\scriptscriptstyle\rm  LQG}$. Following 
\cite{Jaynes}, $\tilde\rho$ can be obtained as the distribution that maximizes Shannon entropy, with the constraint given by available measurements. This gives the density matrix
\be \label{penultima}
       \tilde\rho=Z^{-1}\ e^{-\mu H'},
\ee
which maximizes the entropy
\be
       S=-\tr_{\!\scriptscriptstyle\rm  LQG}[\tilde\rho \log\tilde\rho].
\ee
Here $Z=\tr_{\!\scriptscriptstyle\rm  LQG}[e^{-\mu H'}]$ and $\mu$ is determined by $\bar E$ by $\bar E=\tr_{\!\scriptscriptstyle\rm LQG}[ H'\tilde\rho]=-d(\ln Z)/d\mu$.  Therefore the relation between this gravitational entropy and the Hamiltonian operator is 
\be\label{ultima}
       S=\log Z + \mu \tr_{\!\scriptscriptstyle\rm  LQG}[H'\ e^{-\mu H'}].
\ee
A part from the prefactors, and the exponent instead as the logarithm, the key point here is that the trace and the operator $H'$ are in ${\cal H}_{\scriptscriptstyle\rm  LQG}$ and not in $\H_{\scriptscriptstyle\rm P}$.

In particular, we can take $\psi$ to be the state where the particle is concentrated on a single node $n$.  This give the entropy of a spin network $s$
\be
 \label{penultima}
       \tilde\rho(s)=Z^{-1}\ e^{-\mu d_s},
\ee
where $d_s$ is the degree of a single node.  A distribution that is probably more physically motivated, since it avoids singling out a node, can be defined by taking instead the average of the degrees over all the nodes of the spin network. This corresponds to the entropy defined by the coarse graining obtained by measuring the geometry with a particle concentrated on one node, but without knowing which one. Then 
\be
 \label{penultima}
       \tilde\rho(s)=Z^{-1}\ e^{-\mu \bar d_s},
\ee
where
\be
\bar d_s(\mu) =  \f{ \sum_n^N d_n}{N}= \f{2\lin}{N}.
\ee
The partition function can be easily calculated \cite{francis}, and gives
\be
Z(\mu)=\sum_s e^{-\mu \bar d_s} 
=
\sum_{\lin=0}^L  \left( \!
\begin{array}{c}
L \\[1pt]
\lin \end{array} 
\!
\right)   e^{-\mu \f{2\lin}{N}} 
=
\left( 1 + e^{-\mu\f2N}  \right)^L
\ee
This is a very simple possible distribution over spinnetworks. Its properties will be explored elsewhere. 

\section{Conclusions}

In the first part of the paper we have constructed the Hamiltonian operator of a quantum particle that moves slowly on a quantum gravitational field.  The operator, given in \eqref{den} has a simple and appealing form. Perhaps this construction can be used to explore more extensively the general form of effective gravity-matter couplings in quantum gravity.  A key result of this part of the paper is that the quantum kinematics and the dynamics of the particle cannot be constructed in a way disjoint form the geometry: the particle wave function turns out to have support on the nodes of the spin network, and its scalar product depend on the volume, namely on the spin network itself.  This result (see also \cite{Ashtekar:2009mb}) can perhaps be relevant for understanding the dynamics of matter in quantum gravity. 

In the second part of the paper we have observed that (under a suitable simplifying assumption) this operator is the BGS density matrix of a graph related to the state of gravity.   We have asked if the corresponding BGS entropy can be interpreted as a genuine physical entropy, related to a probability of finding the state. The structure of the theory is extremely similar to that of statistical mechanics: a trace in the full state space of gravity determines an operator on a smaller space, which has all the mathematical properties of a density matrix and codes the spectral information about the geometry which is available to matter.  

In spite of the close analogy, this mathematical density matrix does not seem interpretable as a physical density matrix, since it does not have a precise probabilistic interpretation. Rather, the density matrix and entropy of the gravitational field determined by a matter spectral measurement are given by \eqref{penultima} and \eqref{ultima}, respectively. Our analysis is incomplete since it disregards the dynamics, namely the restriction on the physical states imposed by the Hamiltonian constraint; but we do not see how this could improve the situation. Therefore, as far as we can see, the BGS entropy of a spin network state does not appear to have a direct probabilistic interpretation.  But the analogy remains tantalizing. 
On the other hand, the Passerini-Severini result shows that the BGS entropy is a useful measure of the regularity of physical space. 

Finally, we have considered a tentative definition of a probability distribution on the space of the spin networks. This is determined by maximizing the Shannon entropy, conditioned to the measurement of the geometry made by measuring the energy of a particle of known position.  The properties of the resulting distribution will be studied elsewhere. 

In closure, we would like to compare our discussion with other attempts to define gravitational entropy.  However, we have found remarkably little on the subject.  The entropy of matter on a given geometry is well understood and discussed everywhere in the literature; the entropy to be associated with a black hole is of course largely discussed as well;  but the entropy of the gravitational field itself (in the full nonperturbative theory) is a notion that, as far we can see, is almost entirely ignored in the literature.  The only discussion we are aware of is the one given by one of us in \cite{Rovelli:1993a,Rovelli:1993b}. See these papers for a more detailed discussion. (See also \cite{Connes:1994hv} and \cite{Smolin:1981jt,Smolin:1982jt} on related subjects.) The statistical mechanics of a generally covariant field theory (which is to say: of our true world) is a chapter of physics still to be built.

%\vspace{.1em}{\ }
\pagebreak
%\centerline{-----}
%\vspace{.1em}{\ }
\subparagraph{Acknowledgments}
Thanks to Filippo Passerini, Alejandro Perez, Matteo Smerlak and Simone Speziale for several useful exchanges.  We thank an anonymous referee for his/her extremely valuable and helpful comments. 
FV acknowledges the support of the Fondazione Angelo Della Riccia. 

\vspace{14mm}{\ }

\section*{APPENDIX}

We give here a naive derivation of the form \eqref{den} of the Hamiltonian constraint. The energy operator of Schr\"odinger wave mechanics on flat space is $H=-\f{\hbar^2}{2m}\partial^a\partial_a$ and its matrix elements are
\be
\bk\psi H \phi = \f{\hbar^2}{2m}\int \dd x\  \overline{\partial^a\psi}\;{\partial_a\phi}.
\ee
Let us approximate this integral by a Riemann sum over the points of a cubic lattice of lattice space $L$, and approximate the derivative by a difference. We have then
\ba
\bk\psi H \phi = \f{\hbar^2}{2m} \ \sum_n\ \ \sum_a \ L^3&
\nn
& \hspace{-22mm}\times \ { \f{ \overline{\psi(x_{n+\hat a})-\psi(x_n)}}{L}
\ 
\f{{\phi(x_{n+\hat a})-\phi(x_n)}}{L}}
\ea
The first sum is over the nodes of the lattice, and the second over the three directions. Summing over these two variables is the same as summing over the links $\lin$ of the lattice, obtaining 
\ba
\bk\psi H \phi = &\!\!\f{\hbar^2}{2m}\ \sum_{\lin} \ L
\left({\overline{\psi(\lin_f)}-\psi(\lin_i)}\right)
\left({{\phi(\lin_f)}-\phi(\lin_i)}\right)\!.
&\nonumber
\ea
This can be rewritten in the form 
\ba
\small\bk\psi H \phi = &\!\!\f{\hbar^2}{2m}\ \sum_{\lin} \ A^2_{\lin}
\left(\overline{\f{\psi(\lin_f)}{V_{\lin_f}}-\f{\psi(\lin_i)}{V_{\lin_i}}}\right)
\left(\f{\phi(\lin_f)}{V_{\lin_f}}-\f{\phi(\lin_i)}{V_{\lin_i}}\right)
&\nonumber\ea
where $A_{\lin}=L^2$ and $V_{n}=L^3$ are the area of the square separating two lattice cells (of the dual lattice), which is cut by the link $\lin$ and the volume of the cell around the node $n$.
Now, let us imagine that instead of being on flat space the particle is on a curved space.  Then the area and volumes $A_{\lin}$ and $V_{n}$ will become non trivial functions of ${\lin}$ and ${n}$.  Finally, in quantum gravity these quantities will have discrete eigenvalues \cite{Rovelli:1994ge, Ashtekar:1996eg}, giving directly \eqref{den}.

\end{document}